\documentclass[onecolumn,12pt,superscriptaddress]{revtex4-1}
\usepackage{amsmath}
\usepackage{graphicx}
\usepackage{float}
\usepackage{color}
\usepackage{citesort}
\usepackage{ulem}

\begin{document}
	
\title{Heat transport in an optical lattice via Markovian feedback control}

\author{Ling-Na Wu}
\email[]{lingna.wu@tu-berlin.de}
%\address{Institut f\"ur Theoretische Physik, Technische Universit\"at Berlin, Hardenbergstra\ss e 36, Berlin 10623, Germany}
%\ead{lingna.wu@tu-berlin.de}

\author{Andr\'e Eckardt}
\email[]{eckardt@tu-berlin.de}
\address{Institut f\"ur Theoretische Physik, Technische Universit\"at Berlin, Hardenbergstra\ss e 36, Berlin 10623, Germany}
%\ead{eckardt@tu-berlin.de}

\graphicspath{{figures/}}

%\date{\today}

\begin{abstract}
Ultracold atoms offer a unique opportunity to study many-body physics in a clean and well-controlled environment. However, the isolated nature of quantum gases makes it difficult to study transport properties of the system, which are among the key observables in condensed matter physics. 
In this work, we employ Markovian feedback control to synthesize two effective thermal baths that couple to the boundaries of a one-dimensional Bose-Hubbard chain. This allows for the realization of a heat-current-carrying state. We investigate the steady-state heat current, including its scaling with system size and its response to disorder. In order to study large systems, we use semi-classical Monte-Carlo simulation and kinetic theory. The numerical results from both approaches 
show, as expected, that for non- and weakly interacting systems with and without disorder one finds the same scaling of the heat current with respect to the system size as it is found for systems coupled to thermal baths. Finally, we propose and test a scheme for measuring the energy flow. Thus, we provide a route for the quantum simulation of heat-current-carrying steady states of matter in atomic quantum gases. 
%show that the current for a system with a fixed particle number exhibits diffusive behavior, i.e., it decays with the lattice size as $M^{-1}$. While a system with a fixed filling factor exhibits ballistic transport. Interactions do not change the current's scaling with system size when the particle number is fixed. In the presence of disorder, the current for a system with a fixed filling factor is found to decay exponentially with the system size. These behaviors are similar to those when the system is coupled to real thermal baths. Our findings can  be tested using available experimental techniques.
\end{abstract}
%\submitto{\NJP}
\maketitle
\section{Introduction}
%\textit{Introduction.}---

Transport plays a crucial role in understanding states
of matter in and out of equilibrium. The study of transport properties in real materials is always influenced by the effect of impurities, lattice defects and phonons. Ultracold atoms, in turn, offer to study transport also in systems, which are isolated from their environment and free of impurities or disorder, unless these properties are engineered on purpose in a controlled fashion. Moreover, they are highly tunable and can be manipulated and probed on their (large) intrinsic time and lengths scales~\cite{RevModPhys.80.885}. This makes them promising quantum simulators also for transport. 
%Ultracold atomic quantum gases are free of impurity or disorder, and are well isolated from the environment. Moreover, the relevant
%length and energy scales differ by
%orders of magnitude from those of an electron gas in a solid.
%These properties make quantum gases a unique platform for quantum simulation of condensed matter models. 
%The measurement of transport constitutes a key element in the toolbox. 
However, the isolated nature of quantum gases prevents a
direct connection of the system, e.g., to leads or extended thermal baths of different temperature. 
To investigate the transport properties of quantum gases, a variety
of approaches have been exploited. 
For instance, particle transport has been 
investigated by
observing the response of the system to variations
of the external potential via measuring the density distribution~\cite{PhysRevLett.122.153602,Brown2019}, the
quasimomentum distribution~\cite{Pasienski2010,PhysRevLett.114.083002}, monitoring the center of mass motion~\cite{PhysRevLett.92.160601,PhysRevLett.93.120401,PhysRevLett.94.120403,PhysRevLett.99.220601}, and
expansion dynamics~\cite{PhysRevLett.95.170409,PhysRevLett.95.170410,PhysRevLett.104.220602,Roati2008,Billy2008,Deissler2010,Kondov2011,Jendrzejewski2012,Choi2016,PhysRevLett.110.205301,Schneider2012}, or by studying 
mass flow through optically structured mesoscopic
devices~\cite{Krinner2015,Krinner2016,Krinner_2017,PhysRevX.8.011053}. Spin transport  was studied by introducing spin
inhomogeneities followed by monitoring the spin
evolution~\cite{sommer2011universal,Fukuhara2013,Valtolina2017,Nichols2019} or investigating the decoherence of spin texture~\cite{Koschorreck2013,PhysRevLett.113.147205,PhysRevLett.118.130405,Bardon2014,PhysRevLett.114.015301,Jepsen2020}.
And heat transport was investigated by locally heating the system, after which the equilibration is studied by monitoring the temperature bias~\cite{PhysRevLett.103.095301} or particle imbalance~\cite{Brantut2013,PhysRevX.11.021034}. However, in all these experiments, transport occurs as a transient phenomenon only. 
%Patel2020

In this work, we employ Markovian feedback control~\cite{PhysRevA.49.2133} to engineer two effective thermal baths that are coupled to a one-dimensional Bose-Hubbard chain. This allows for the realization of a heat-current-carrying steady state.
As a measurement-based approach, Markovian feedback control continuously adds a signal-proportional feedback term to the system Hamiltonian. The dynamics of the system is then described by a feedback-modified Lindblad master equation~(ME)~\cite{PhysRevA.49.2133}. By properly choosing the measurement and feedback operators, the system dynamics can be steered towards a desired target state. The Markovian feedback method has been applied to various
control problems, including the stabilization of arbitrary one-qubit quantum states~\cite{PhysRevA.64.063810,PhysRevLett.117.060502},
the manipulation of quantum entanglement between two qubits~\cite{PhysRevA.71.042309,PhysRevA.76.010301,PhysRevA.78.012334,Wang2010}
as well as optical and spin squeezing~\cite{PhysRevA.50.4253,PhysRevA.65.061801,2021arXiv210202719B}. In our previous works, we have shown that Markovian feedback control can be used to cool a bosonic quantum gas in an optical lattice~\cite{wu2021cooling} and to  engineer a thermal bath~\cite{wu2022FiniteT}. Here, we will generalize such a feedback scheme to engineer two thermal baths at the boundaries of an optical lattice and study the heat transport through the chain induced by it. Different from our previous works~\cite{wu2021cooling,wu2022FiniteT}, this requires to work out a scheme for engineering artificial thermal baths by employing local measurements and feedback  on a few lattice sites only. 

The paper is organized as follows: in Section~\ref{model}, we introduce our model and some basics of Markovian feedback control. This is followed by the description of a two-site feedback scheme in Section~\ref{scheme}, which can be used to engineer a finite-temperature bath. In Section~\ref{transport}, we study the steady-state heat current of the system, including its scaling behavior with system size~(see Section~\ref{sec:system-size-scaling}) and its response to disorder~(see Section~\ref{sec:disorder}). The experimental implementation of our scheme is discussed in Section~\ref{sec:EI}, including the measurement of heat current by measuring single-particle density matrix. A summary of the main results is presented in Section~\ref{sec:conclusion} to conclude.

\section{Model and Markovian feedback scheme}\label{model}
%\textit{Model.}---
%\deleted{We describe a system of $N$ bosons with open boundary conditions using the one-dimensional}
The system under consideration is a one-dimensional optical lattice with $N$ interacting bosonic atoms, which can be described by the Bose-Hubbard model,
\begin{equation}\label{BH}
	H = -J\sum\limits_{l=1}^{M-1}(a_l^\dag a_{l+1} + a_{l+1}^\dag a_l) + \frac{U}{2}\sum\limits_{l=1}^{M}n_l (n_l-1) + \sum\limits_{l=1}^{M}{V_l n_l},
\end{equation}
where $a_l$ annihilates a particle on site $l$ and $n_l=a_l^\dag a_l$ counts the particle
number on site $l$, with $\sum\nolimits_{l}n_l=N$.
The first term in (1) describes tunneling between neighboring sites with rate $J$, the second term
denotes on-site interactions with strength $U$ and the last term describes an on-site potential. In the following discussion, $V_l=0$ unless stated otherwise.

Let us consider a homodyne measurement of an operator $c$.
The dynamical evolution of the system
is then described by the stochastic master equation~(SME)~\cite{wiseman2009quantum}~($\hbar=1$ hereafter),
%$d\rho_c = -i[H,\rho_c] dt + {\cal D}[c]\rho_c dt + {\cal H}[c]\rho_c dW$,
\begin{equation}\label{SME}
	d\rho_c = -i[H,\rho_c] dt + {\cal D}[c]\rho_c dt + {\cal H}[c]\rho_c dW, \notag
\end{equation}
with ${\cal H}[c]\rho := c\rho + \rho c^\dag - {\rm Tr}[(c+c^\dag)\rho]\rho$
and ${\cal D}[c]\rho := c\rho c^\dag - \frac{1}{2}(c^\dag c \rho + \rho c^\dag c)$.
Here $\rho_c$ denotes the quantum state conditioned on the measurement result,
$I_{\rm hom} = {\rm Tr}[(c+c^\dag)\rho] + \xi(t)$, with $\xi(t)=dW/dt$ and $dW$
being the standard Wiener increment with mean zero and variance $dt$.
The quantum backaction of a weak measurement can be used for tailoring the system's dynamics and to
prepare target states. 
%For instance, a quantum nondemolition measurement of light allows for the preparation of different types of atom-number squeezed and macroscopic superposition states~\cite{PhysRevLett.102.020403,PhysRevLett.114.113604}.
While the state generated in this way is conditional due to
the nondeterministic nature of measurement, the introduction of feedback using the information acquired from the
measurements allows to steer the system's dynamics into a desired state.

Here we consider a direct feedback strategy, where a signal-dependent, i.e. conditional, feedback term
$I_{\rm hom} F$ is added to the Hamiltonian. According to the theory of Markovian feedback control~\cite{PhysRevA.49.2133},
the system is then governed by the feedback-modified SME
\begin{eqnarray}\label{sme_fbM}
	d\rho_c &=& -i[H+H_{\rm fb},\rho_c]dt + {\cal D}[A]\rho_c dt +{\cal H}[A]\rho_c dW, \notag
\end{eqnarray}
with operators
\begin{equation}\label{A}
	A = c-iF,  \quad H_{\rm fb}=\frac{1}{2}(c^\dag F+F c).
\end{equation}
By taking the ensemble average of the possible measurement outcomes, we arrive at the feedback-modified ME~\cite{PhysRevA.49.2133}
\begin{equation}\label{me_fbM}
	\dot \rho = -i[H+H_{\rm fb},\rho] + {\cal D}[A]\rho.
\end{equation}
%with $ H_{\rm fb}=\gamma(c^\dag F+ F c)/2$ and $A=c-i F$.
The effect induced by the feedback loop is seen to replace the collapse
operator $c$ by $A$ and to add an extra term $H_{\rm fb}$ to the Hamiltonian. The latter is proportional to measurement strength and thus can be safely neglected for weak measurements.

\section{Two-site feedback scheme}\label{scheme}
Before approaching the scenario relevant for heat transport, where the feedback is mimicking two thermal baths of different temperature at both ends of the system, let us first investigate how the {\it {local}} coupling to a single bath can be realized. Previously, we considered already the engineering of a thermal bath using measurement and feedback operators acting globally on all sites of a lattice~\cite{wu2022FiniteT}. In contrast, we now consider the following two-site measurement and feedback operator
\begin{equation}\label{two-site-v}
	\begin{aligned}
	c_l &= \sqrt{\gamma}(x_l n_l - {x_l^{-1}}n_{l+1}),  \quad
	F_l &= -i  \lambda \sqrt{\gamma}  (a_{l}^\dag a_{l+1} - a_{l+1}^\dag a_l),
	\end{aligned}
\end{equation}
where $\gamma$ is the measurement strength, $\lambda$ is a free parameter to be determined, $x_l=g_{l+1}/g_l$ and $g_l$ are the coefficients of the single-particle ground state, i.e., $|g\rangle = \sum\nolimits_{l}{g_l |l\rangle}$. The feedback-modified collapse operator then reads 
\begin{equation}
A_l = c_l-iF_l.
\end{equation}
Note that the sites $l$ and $l+1$ where to perform the measurement and feedback can be any neighboring two sites on the lattice. 

\begin{figure}
	\centering
	\includegraphics[width=0.99\columnwidth]{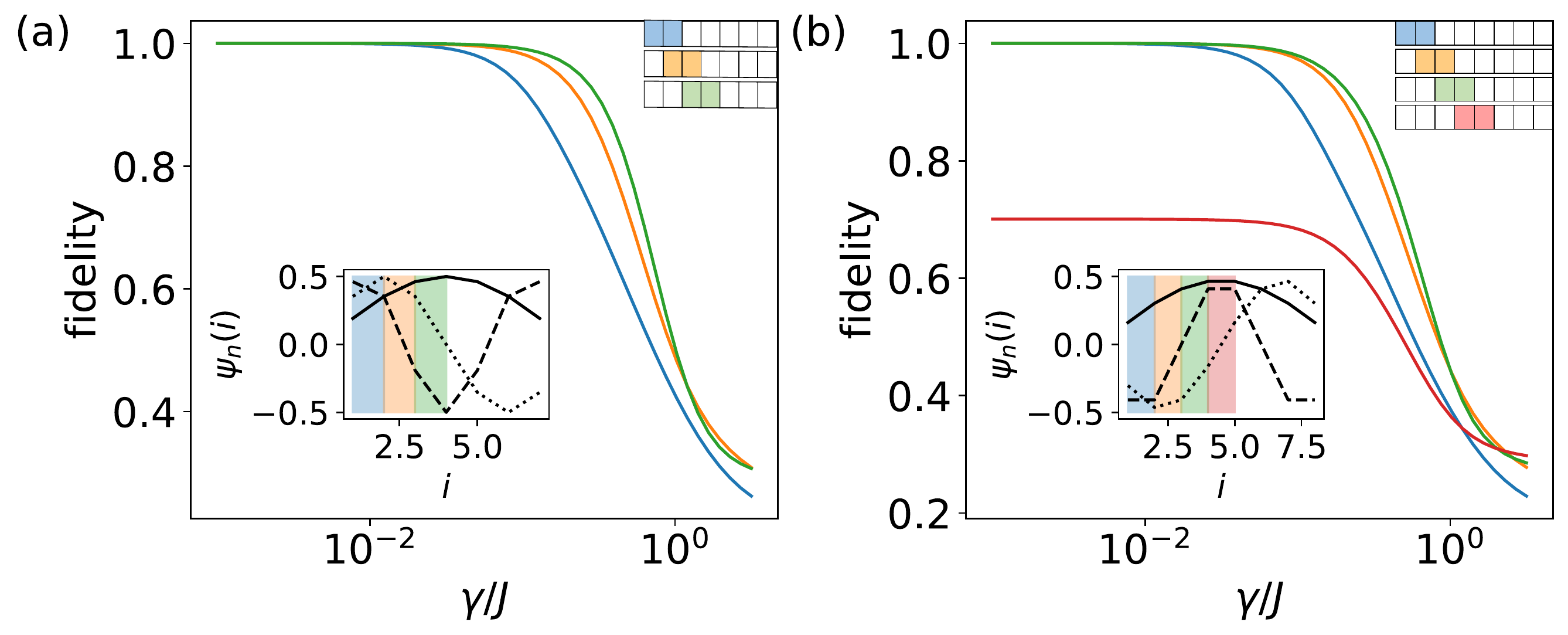}
	\caption{%{Preparing the ground state.}
		The fidelity~\eqref{fid} between the steady state of the ME~\eqref{me_fbM} for our two-site feedback scheme~\eqref{two-site-v} and the ground state of the system~\eqref{BH} as a function of the measurement strength $\gamma$ for two different lattice sizes with $N=2$ non-interacting particles: (a) $M=7$ and (b) $M=8$. Different colored curves correspond to schemes performing at different sites as indicated by the sketch at the upper right corner with the corresponding colors. As expected, for weak measurements $\gamma \ll J$, the fidelity approaches $1$. The only case where the feedback-controlled system does not settle down to the ground state [see the red curve in (b)] for weak measurements is due to the fact that the ground state is not the unique dark state of the collapse operator. From the inset, which shows the first three eigenstates of the system, one can see that the first excited state for $M=8$ in (b) as denoted by the dashed line has the same wavefunction values as the ground state~(see solid line) at the feedback-controlled sites~(site $4$ and site $5$), and thus is also a dark state of the collapse operator.}\label{comparev}
\end{figure}

%Here we show that for non-interacting particles, with $\lambda=1$, $A_l |g\rangle^{\otimes N} = 0$, where $|g\rangle^{\otimes N}$ denotes the ground state of the system, with all particles occupying the single-particle ground state, $|{g}\rangle$. 

For $N$ non-interacting particles, with $\lambda=1$, one can show that $A_l |g\rangle^{\otimes N} = 0$, where $|g\rangle^{\otimes N}$ denotes the ground state of the system, with all particles occupying the single-particle ground state, $|{g}\rangle$. 
%Firstly, we show that this holds for a single particle. In this case, 
It is easy to check it for the single-particle problem, where
the collapse operator reduces to 
\begin{equation}
	A_l = \frac{g_{l+1}}{g_l}{|l\rangle\langle l|} - \frac{g_{l}}{g_{l+1}}{|l+1\rangle\langle l+1|} - (|l\rangle\langle l+1|-|l+1\rangle\langle l|). \notag
\end{equation}
Applying it to the ground state $|g\rangle$, one gets
\begin{equation}
	A_l|g\rangle = g_{l+1}|l\rangle - g_l|l+1\rangle-g_{l+1}|l\rangle+g_l|l+1\rangle=0. \notag
\end{equation}
When there are no interactions between the particles, the multi-particle problem is equivalent to the single-particle problem.
%For $N$ non-interacting particles, it is easy to show that with $\lambda=1$,  $A_l |g\rangle^{\otimes N} = 0$~(see~\ref{append:prove}), where $|g\rangle^{\otimes N}$ denotes the ground state of the system, with all particles occupying the single-particle ground state, $|{g}\rangle$. 
Namely, the ground state of the system is a dark state of the collapse operator $A_l$. Assuming weak measurements with strength $\gamma \ll J$, where the impact of the additional term in the Hamiltonian $H_{\rm bf} \propto \gamma$ is negligible, the dissipative dynamics will then
drive the system towards the ground state (if it is the unique dark state of the collapse operator).  %This approach is similar to the cooling scheme proposed in Ref.~\cite{wu2021cooling}, while the measurement and feedback operators for the latter spread over the whole lattice.   

\begin{figure}
	\centering
	%\includegraphics[width=0.99\columnwidth]{figure/two_site_current_v.pdf}
	\includegraphics[width=0.99\columnwidth]{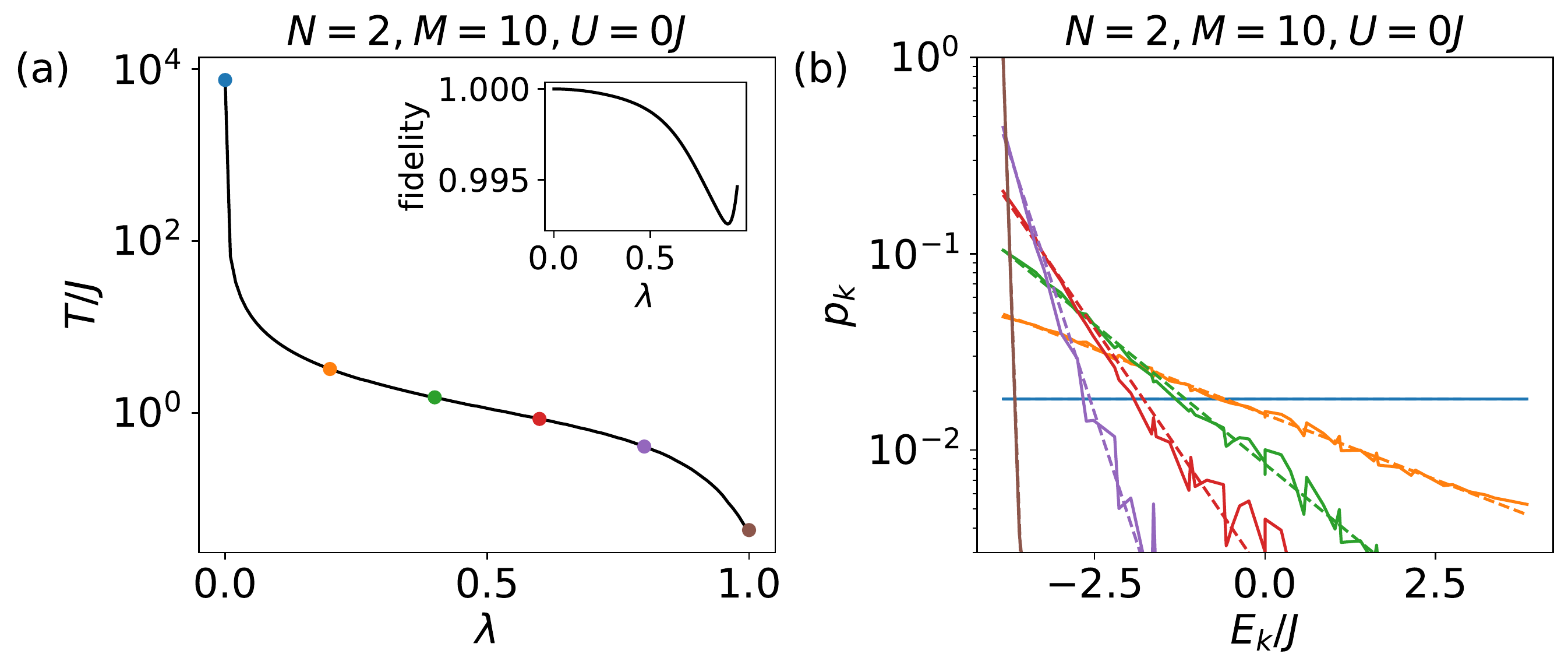}
	\includegraphics[width=0.99\columnwidth]{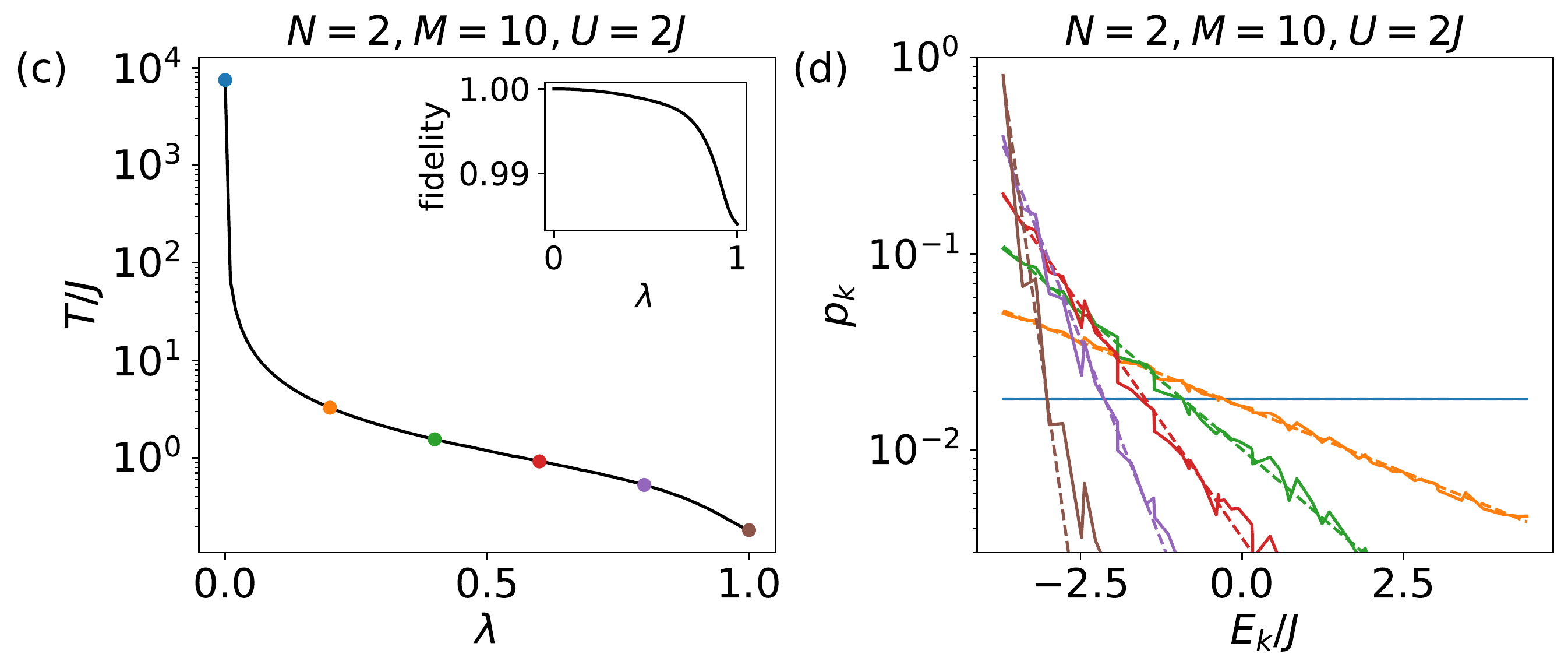}
	\caption{{Engieering a finite-temperature bath using two-site feedback scheme~\eqref{two-site-v}.} (a, c)~The effective temperature of the feedback-synthesized bath as a function of the feedback strength $\lambda$. The inset shows the fidelity between the steady state and the corresponding effective thermal state. (b, d)~The distribution of the steady state on the eigenbasis for various $\lambda$ marked in (a, c). $k$ is the label of the $k$th eigenstate, with energy in ascending order. Dashed lines are the corresponding thermal distributions. Parameters are  $V_l=0$, $\gamma=0.01J$.}\label{pkt}
\end{figure}

In Fig.~\ref{comparev}, we show
the fidelity between the steady state of the ME~\eqref{me_fbM} for our two-site feedback scheme~\eqref{two-site-v}, $\rho_{\rm ss}$, and the ground state of the system~\eqref{BH}, $|G\rangle$, i.e.,
\begin{equation}\label{fid}
	f = \sqrt{\langle G| \rho_{\rm ss} |G \rangle},
\end{equation}
as a function of the measurement strength $\gamma$ for two different lattice sizes with $N=2$ non-interacting particles: (a) $M=7$ and (b) $M=8$. Different colored curves correspond to schemes performing at different sites as indicated by the sketch at the upper right corner with the corresponding colors. As expected, for weak measurements $\gamma \ll J$, the fidelity approaches $1$. The only case where the feedback-controlled system does not settle down to the ground state [see the red curve in (b)] for weak measurements is due to the fact that the ground state is not the unique dark state of the collapse operator. Namely, from the inset, which shows the first three eigenstates of the system, one can see that the first excited state for $M=8$ in (b) as denoted by the dashed line has the same wavefunction values as the ground state~(see solid line) at the feedback-controlled sites~(site $4$ and site $5$), and thus is also a dark state of the collapse operator.

For $0<\lambda<1$, the proposed scheme can be used to engineer a finite-temperature bath. In Fig.~\ref{pkt}, we show the effective temperature of the feedback-synthesized bath as a function of the feedback strength $\lambda$ for (a) non-interacting and (c) interacting systems. The inset shows the fidelity between the steady state and the corresponding effective thermal state, which is close to $1$ over the whole parameter regime. As a second measure, we compare the probability distribution in the eigenstate basis for the steady states at various $\lambda$~(solid lines) marked in Fig.~\ref{pkt}(a, c) to the corresponding thermal states~(dashed lines) in Fig.~\ref{pkt}(b, d). They are found to agree with each other very well.

%\newpage

\section{Heat transport}\label{transport}
\subsection{Model}

\begin{figure}
	\centering
	\includegraphics[width=0.6\columnwidth]{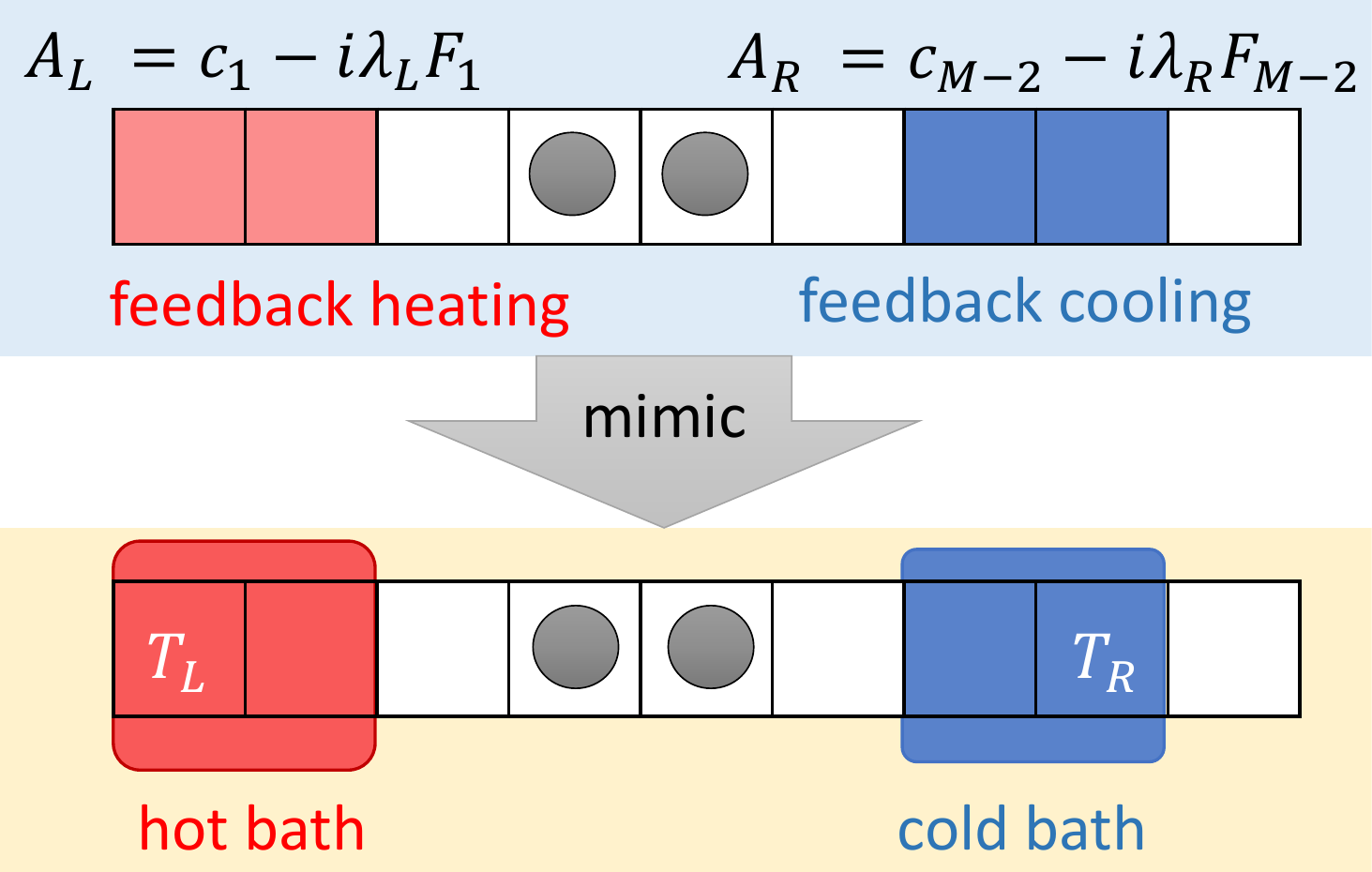}
	\caption{{A sketch of our feedback scheme for realizing a heat-current-carrying state.} A one-dimensional optical lattice is cooled and heated at both ends via feedback control, as shown in the upper panel in blue background. This mimics the effect of coupling the system locally to a hot  bath and a cold bath~(see lower panel in yellow background), and thus allows us to study the heat transport property of the system. 
	}\label{fb_model}
\end{figure}

We use the two-site feedback scheme to heat up the system on one side and cool it down on the other side, see the sketch in Fig.~\ref{fb_model}. Note that the feedback-controlled sites for the bath on the right hand side of the chain are chosen differently from the left hand site, not to be the outermost sites, otherwise the system will be effectively coupled to one thermal bath at the average temperature of the two synthesized baths~\cite{PhysRevE.92.062119}. {{The two measurement operations are assumed to be independent of each other~(see Section~\ref{sec:EI} for the experimental implementation), giving rise to uncorrelated signals.}} The dynamics of the system is described by the feedback master equation,
\begin{equation}\label{fb_me}
	\dot \rho = {\cal L}\rho=-i[H,\rho] + {\cal L}_L \rho + {\cal L}_{R}\rho,
\end{equation}
where
\begin{equation}
	{\cal L}_\mu \rho = -i[H_{{\rm fb}}^{\mu}, \rho] + {\cal D}[A_\mu](\rho), \quad (\mu=L, R)
\end{equation}
describes the impact of the feedback control on the $\mu$ side of the chain
with $H_{{\rm fb}}^{L} \equiv H_{{\rm fb}}^{(1)}$, $H_{{\rm fb}}^{R} \equiv H_{{\rm fb}}^{(M-2)}$, $H_{\rm fb}^{(l)} = (c_{l} F_{l} + F_{l}c_{l})/2$ and
\begin{equation}
	A_L \equiv A_1 = c_1 - i\lambda_L F_1, \quad A_R \equiv A_{M-2} = c_{M-2} - i\lambda_R F_{M-2}.
\end{equation}
%\begin{equation}
%	{\cal D}[A_i](\rho) = A_i\rho A_i^{\dag} - \frac{1}{2}A_i^{\dag}A_i\rho - \frac{1}{2}\rho A_i^{\dag} A_i,
%\end{equation}
%and 
%\begin{equation}
%	A_i = c_i-i{\lambda_i}F_i.
%\end{equation}
%Here,
%\begin{eqnarray}\label{two-site-v}
%	c_i &=& \sqrt{\gamma}(x n_i - {x^{-1}}n_{i+1}), \notag\\
%	F_i &=& -i  \sqrt{\gamma}  (a_{i}^\dag a_{i+1} - a_{i+1}^\dag a_i),
%\end{eqnarray}
%where $x=g_{i+1}/g_i$, and $g_i$ is the coefficient of the ground state, i.e., $|G\rangle = \sum\nolimits_i{g_i |i\rangle}$. 
Cooling is realized by setting $\lambda=1$, corresonding to a zero-temperature bath, and heating by setting $0<\lambda<1$, corresponding to finite positive temperatures.
%In the following simulations, we keep $\lambda_{M-1}=1$, i.e., cooling on the right side, and $0<\lambda_1<1$, heating on the left side.

\subsection{Heat current}
The steady state of the system is a heat-current-carrying state.
The heat current ${\cal J}$ is calculated from the continuity equation for energy,
\begin{equation}\label{continue_eq}
	\frac{d\langle H \rangle}{dt} = {\cal J}_L + {\cal J}_R. 
\end{equation}
It follows from the ME~\eqref{fb_me} that
\begin{equation}
	\frac{d\langle H \rangle}{dt} = {\rm Tr}\left\{H{\cal L}_L\rho\right\} + {\rm Tr}\left\{H{\cal L}_{R}\rho\right\}. \notag
\end{equation}
So we have
\begin{equation}
	{\cal J}_L = {\rm Tr}\left\{H{\cal L}_L\rho\right\}, \quad {\cal J}_R = {\rm Tr}\left\{H{\cal L}_{R}\rho\right\}.
\end{equation}
In the steady state, $d\langle H\rangle/dt = 0$.
Thus, the steady-state heat current is
\begin{equation}
	{\cal J}_{\rm ss} = {\cal J}_L = -{\cal J}_R = {\rm Tr}\left\{H{\cal L}_L\rho_{\rm ss}\right\} = -{\rm Tr}\left\{H{\cal L}_{R}\rho_{\rm ss}\right\}.
\end{equation}
%where $\rho_{\rm ss}$ is the steady state.

\begin{figure}
	\centering
	\includegraphics[width=0.9\columnwidth]{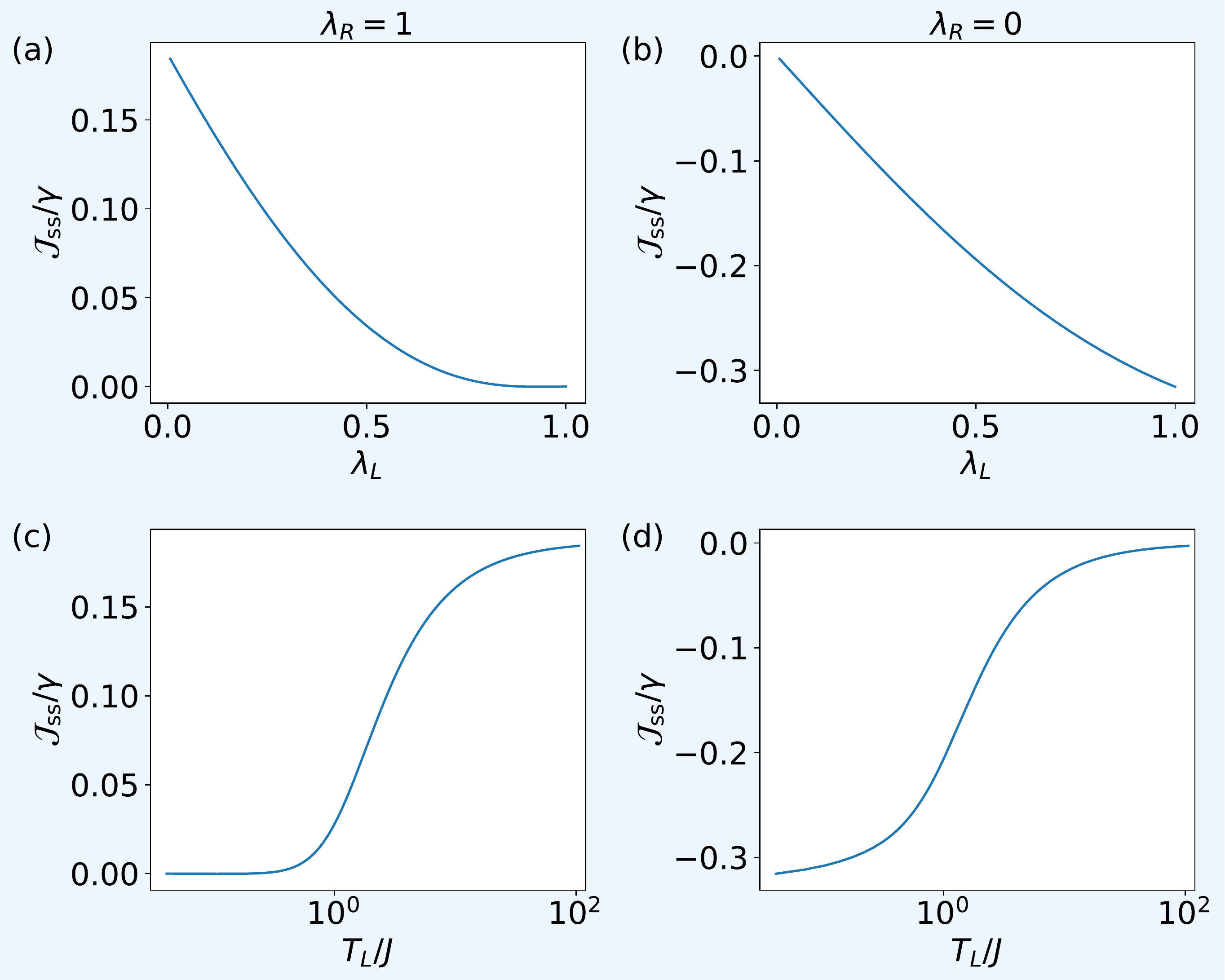}
	\includegraphics[width=0.9\columnwidth]{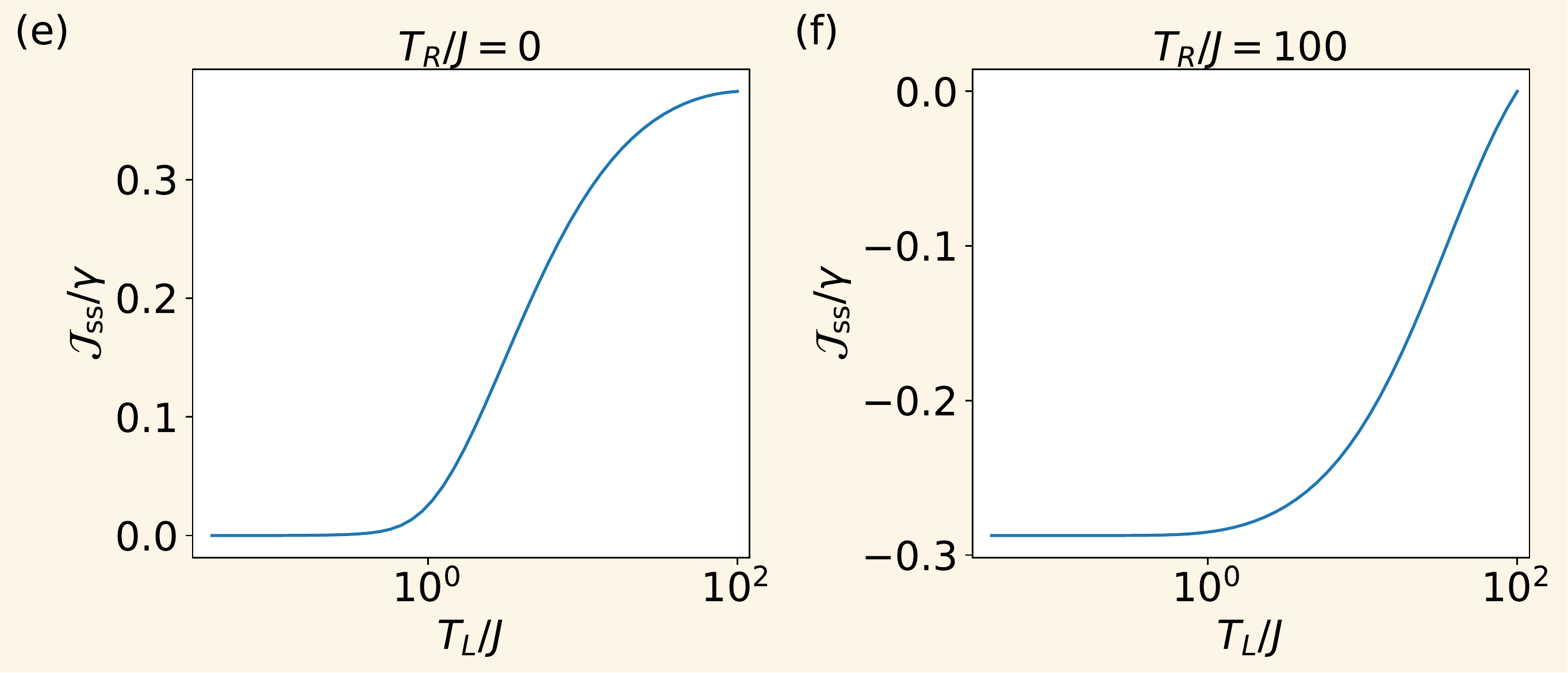}
	\caption{The panels in blue background (a-d) are the results for our scheme. The steady-state heat current is shown as a function of (a, b) the feedback strength $\lambda_L$ and (c, d) the effective temperature $T_L$ of the left bath. For the right bath, (a, c) $\lambda_R=1$, which corresponds to a zero-temperature bath; (b, d) $\lambda_R=0$, which corresponds to an  infinite-temperature bath. In both cases, the heat current increases with the temperature imbalance between the two baths. Similar behavior is observed when the system is coupled to real thermal bath, as shown in the panels (e, f) in yellow background.
	Parameters are $N=1$, $M=10$, $V_l=0$, $\gamma=0.01J$. 
	}\label{current}
\end{figure}

Figure~\ref{current} shows the steady-state heat current of a system with $N=1$ particle in a lattice with $M=10$ sites as a function of the feedback strength $\lambda_L$~[see (a) and (b)] and the effective temperature $T_L$~[see (c) and (d)] of the left bath. 
For the right bath, in (a, c) $\lambda_R=1$, which corresponds to a zero-temperature bath~($T_R=0$); in (b, d) $\lambda_R=0$, which corresponds to an  infinite-temperature bath~($T_R=\infty$). As expected, the heat current in both cases increases with the temperature imbalance between the two baths. Similar behavior is observed when the system is coupled to real thermal bath, as shown in Fig.~\ref{current}(e, f). Since the heat current in non-equilibrium is not solely determined by the temperature of the baths, we do not expect exactly the same behavior for our scheme and the thermal bath case.
In the following discussion, we will focus on the case with $\lambda_R=1$ for our scheme.

\subsection{System-size scaling}\label{sec:system-size-scaling}
We are interested in the scaling of the steady-state heat current with system size. To study this property, we have to deal with large systems, which are not accessible by  exact diagonalization. For a system with $N$ particles and $M$ sites, the dimension of the Hilbert space is $D=(N+M-1)!/N!/(M-1)!$, which means the Liouvillian superoperator ${\cal L}$ is a $D^2$ by $D^2$ matrix. For instance, for $M=4$ and $N=8$, $D=330$, the Liouvillian superoperator will be a $108900$ by $108900$ matrix. This simple example shows that it is hard to treat large systems by using the exact diagonalization approach. In order to circumvent this problem, we resort to two different approaches: kinetic theory and semi-classical Monte Carlo simulation, as described in the following. 

For the non-interacting case, the system Hamiltonian reads $$H=\sum\limits_k {\epsilon_k n_k },$$ with single-particle eigenenergy $\epsilon_k =-2J\cos{\frac{k \pi}{M+1}}$ in the absence of on-site potential. Here $n_k=c_k^\dag c_k$ counts the number of particles in the single-particle eigenstate $|k\rangle$, with $c_k^\dag = \sum\nolimits_{l}\langle l|k\rangle a_l^\dag$ being the corresponding creation operator. The continuity equation for energy then reads
\begin{equation}
	\frac{d\langle H \rangle}{dt} = \sum\nolimits_k \epsilon_k \langle \dot n_k \rangle.
\end{equation}
It depends on the time evolution of the mean occupations $\langle n_k\rangle$, which is governed by 
\begin{equation}\label{ni}
	\langle \dot n_k\rangle = \sum\nolimits_q\left[R_{kq} \langle n_q(1+ n_k)\rangle - R_{qk}\langle n_k(1+ n_q)\rangle\right], 
\end{equation}
with $R_{kq}=R_{kq}^L + R_{kq}^R = |\langle k|A_L|q\rangle|^2 + |\langle k|A_{R}|q\rangle|^2$ being the single-particle transfer rate from single-particle eigenstate $q$ to $k$. 
%This set of equation is not closed as the single-particle correlations depend on two-particle correlations, which in turn depend on three-particle correlations, and so on. For instance, the equation of motion for the two-particle correlation takes the form~\cite{PhysRevE.92.062119}
%\begin{equation}
%	\begin{aligned}
%	\frac{d\langle n_k n_q\rangle}{dt} &= \sum\limits_{p}{\left[(\chi_{kp}+\chi_{qp})\langle n_k n_q n_p\rangle + R_{kp}\langle n_q n_p\rangle +R_{qp}\langle n_k n_p\rangle -(R_{pk}+R_{pq})\langle n_k n_q\rangle\right]} \\
%	&+\delta_{kq}\sum\limits_{p}{\left[R_{kp}(\langle n_p\rangle+\langle n_k n_p\rangle)+R_{pk}(\langle n_k\rangle+\langle n_kn_p\rangle)\right]}\\
%	&-R_{qk}(\langle n_k\rangle + \langle n_k n_q\rangle) - R_{kq}(\langle n_q\rangle+\langle n_k n_q\rangle),
%	\end{aligned}
%\end{equation}
%with $\chi_{kq}\equiv R_{kq}-R_{qk}$. Hence, we need to make some approximations to do further calculations.
The steady-state heat current is given by
\begin{equation}\label{Jss}
	{\cal J}_{\rm ss} = {\cal J}_L = \sum\nolimits_k \epsilon_k \langle \dot n_k \rangle_L = - {\cal J}_R = -\sum\nolimits_k \epsilon_k \langle \dot n_k \rangle_R
\end{equation}
with
\begin{equation}\label{dnkL}
	\langle \dot n_k\rangle_\mu = \sum\nolimits_q\left[R_{kq}^\mu \langle n_q (1+ n_k)\rangle_{\rm ss} - R_{qk}^\mu\langle n_k(1+n_q)\rangle_{\rm ss}\right],  \quad (\mu=L, R).
\end{equation}
Here the subscript `ss' of the expectation values denotes the steady-state expectation values, which satisfy
\begin{equation}\label{niss}
	\sum\nolimits_q\left[R_{kq} \langle n_q(1+ n_k)\rangle_{\rm ss} - R_{qk}\langle n_k(1+ n_q)\rangle_{\rm ss}\right]=0. 
\end{equation}
In the following, we describe two approaches to calculate the steady-state expectation values approximately.

\subsubsection{Semi-classical Monte Carlo simulation.}

%In order to check the validity of the mean-field results, we perform a semi-classical Monte-Carlo simulation~\cite{PhysRevE.92.062119}. 
In the semi-classical Monte-Carlo simulation~\cite{PhysRevE.92.062119}, the density matrix is approximated by a mixed superposition of Fock states with respect to single-particle eigenstates $\rho = \sum\nolimits_{\bf n}{p_{\bf n}|{\bf n}\rangle \langle {\bf n}|}$, with ${\bf n}=(n_1, n_2, \ldots, n_M)$, i.e., the off-diagonal elements which decouple with the diagonal elements and decay with time are neglected for weak system-bath coupling~\cite{BRE02}. 
The equations of motion for the Fock-space occupation probabilities $p_{\bf n}$ are then mapped to a random walk in the classical space spanned by the Fock states $|{\bf n}\rangle$~(but not their superpositions). We perform these simulations by using the Gillespie-type algorithm described in Ref.~\cite{PhysRevE.92.062119}. By averaging over the long-time dynamics of many trajectories, we can then compute steady-state expectation values, $\langle {n}_k\rangle_{\rm ss}, \langle {n}_k {n}_q\rangle_{\rm ss}$, etc. The steady-state heat current is then calculated by using Eq.~\eqref{Jss}. This approach gives accurate
results after sufficient statistical sampling. For a given accuracy, the sampling size increases with increasing system sizes.

\subsubsection{Kinetic theory.}
%The above results are obtained through exact diagonalization.
We use kinetic theory to treat large systems where the semi-classical Monte Carlo simulation is computationally expensive.
The set of equations~\eqref{ni} is not closed as the single-particle correlations depend on two-particle correlations, which in turn depend on three-particle correlations, and so on. To get a closed set of equations, we employ the mean-field approximation  $\langle n_k n_q\rangle \approx \langle n_k\rangle \langle n_q \rangle$, which then leads to
\begin{equation}\label{mf}
	\langle \dot n_k\rangle \approx \sum\nolimits_q\left\{R_{kq} \langle n_q\rangle [1+\langle n_k\rangle] - R_{qk}\langle n_k\rangle[1+\langle n_q\rangle]\right\}. \notag
\end{equation}
%The steady-state mean occupations are obtained by $\langle \dot n_k \rangle =0$. 
The steady-state heat current is calculated approximately by using Eq.~\eqref{Jss}
%\begin{equation} \label{Jss_mf}
%	{\cal J}_{\rm ss} \approx \sum\nolimits_k \epsilon_k \langle \dot n_k \rangle_L  + \sum\limits_{k,q} \left(2U\sum\limits_{l}{|\langle l |k\rangle|^2 |\langle l|q\rangle|^2} \right)\langle \dot n_k \rangle_L \langle n_q \rangle_{\rm ss},
%\end{equation}
with
%For the steady state, we have
%\begin{equation}\label{mfss}
%	\sum\nolimits_q\left\{R_{kq} \langle n_q\rangle [1+\langle n_k\rangle] - R_{qk}\langle n_k\rangle[1+\langle n_q\rangle]\right\} = 0.
%\end{equation}
\begin{equation}
	\langle \dot n_k\rangle_L \approx \sum\limits_q\left\{R_{kq}^L \langle n_q\rangle_{\rm ss} [1+\langle n_k\rangle_{\rm ss}] - R_{qk}^L\langle n_k\rangle_{\rm ss}[1+\langle n_q\rangle_{\rm ss}]\right\},
\end{equation}
where the steady-state expectation values are obtained by solving $\langle \dot n_k \rangle =0$, i.e.,
\begin{equation}
\sum\limits_q\left\{R_{kq} \langle n_q\rangle_{\rm ss} [1+\langle n_k\rangle_{\rm ss}] - R_{qk}\langle n_k\rangle_{\rm ss}[1+\langle n_q\rangle_{\rm ss}]\right\} = 0.
\end{equation}

\subsubsection{Results.}
%\subsubsubsection{Non-interacting case}

\begin{figure}
	\centering
	%\includegraphics[width=0.9\columnwidth]{figure/MC_N2_M35_U0.pdf}\\
	%	\includegraphics[width=0.9\columnwidth]{figure/dim_log.pdf}
	%\includegraphics[width=1.\columnwidth]{figure/FB_current_Mvec.pdf}
	\includegraphics[width=0.9\columnwidth]{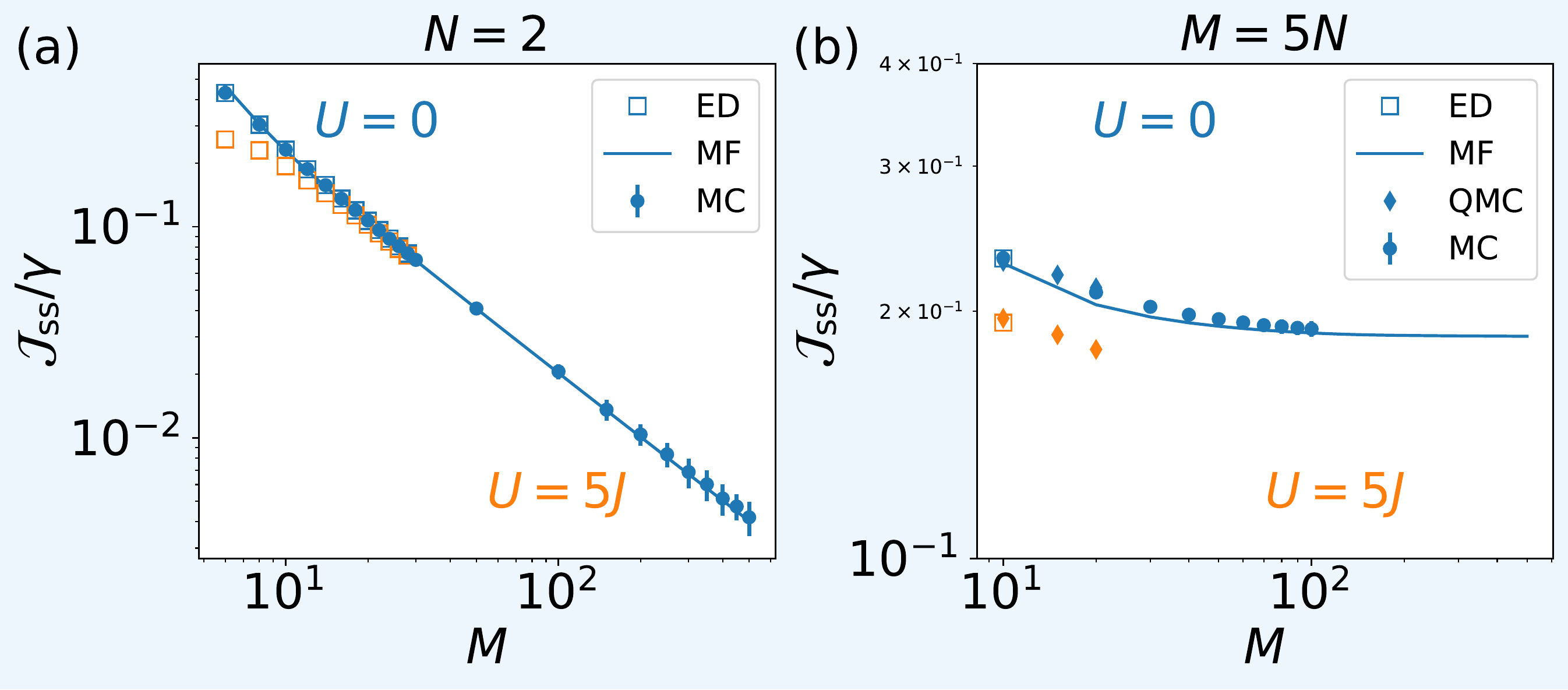}\\
	\includegraphics[width=0.9\columnwidth]{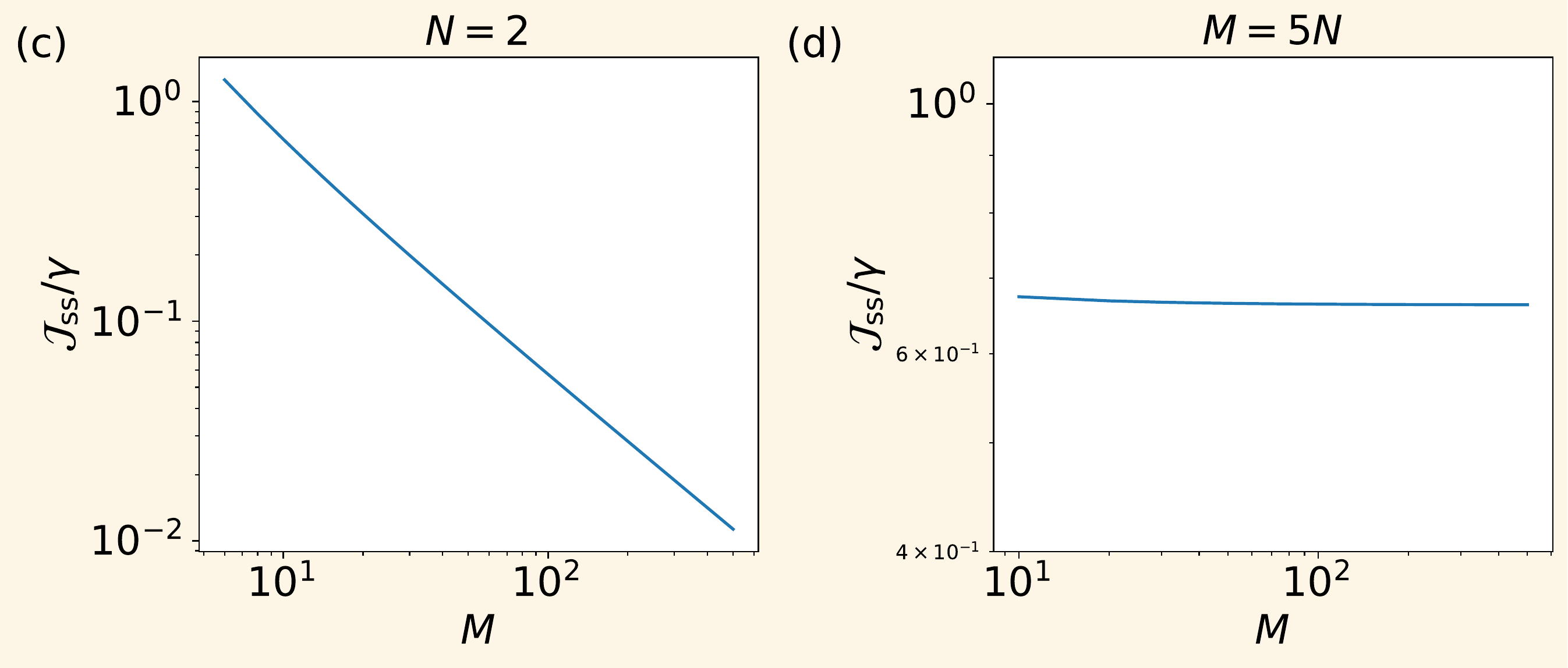}
	\caption{System-size scaling of the steady-state heat current for (a, c) a fixed particle number $N=2$ and (b, d) a fixed filling factor $n=N/M=1/5$. The panels in blue background (a, b) are the results for our scheme. The panels in yellow background (c, d) are the results when the system is coupled to thermal bath.
		The blue~(orange) data denote the results for the non-interacting~(interacting) case. The squares denote the results from exact diagonalization~(ED). The diamonds denote the results from quantum jump Monte-Carlo simulations~(QMC). The bullets denote the results from semi-classical Monte Carlo~(MC) simulation. The solid lines are the results from kinetic theory with mean-field~(MF) approximation. For the MC simulation, the results are obtained by averaging over the long-time~[up to $\gamma t = 10^5$ for (a) and $\gamma t = 10^6$ for (b)] dynamics of $100$ trajectories. The error bar denotes one standard deviation. Parameters are  $V_l=0$, $\gamma=0.01J$, $\lambda_L=0.2$, $\lambda_R=1$, $T_L=10J$, $T_R=0.01J$.
	}\label{MF1}
\end{figure}

Figure~\ref{MF1} shows the system-size dependence of the steady-state heat current. Let us first focus on the non-interacting case, as shown by the blue data. For a fixed particle number $N=2$~[see Fig.~\ref{MF1}(a)], the results from the three approaches, i.e., exact diagonalization~(squares), Monte Carlo simulation~(bullets), and mean-field approximation~(solid lines), agree with each other. The steady-state heat current is found to decrease with the lattice size $M$ as $M^{-1}$, like for diffusive transport~\cite{2021arXiv210414350L}. 
For a fixed filling factor at $N/M=1/5$~[see Fig.~\ref{MF1}(b)], a slight deviation is found between the mean-field results and the Monte Carlo results. %, with the latter showing a better agreement with the exact results for small systems. 
Nevertheless, the results from both approaches show that in this case the current first decreases with system size, but then saturates to a finite value,  independent of the lattice size $M$, corresponding to  ballistic transport~\cite{2021arXiv210414350L}. 
These results are consistent with that when the system is coupled to two thermal baths on its ends, as shown in Fig.~\ref{MF1}(c) and (d).

%\subsubsubsection{Interacting case}
%Now let us turn on interactions. For both fixed particle number~[see Fig.~\ref{MF1}(a) orange data] and fixed filling factor~[see Fig.~\ref{MF1}(b) orange data], the interactions are found to have some impact on the steady-state heat current for small systems. While this effect becomes weaker with increasing system sizes, and thus does not change the system-size scaling behavior of the current. Note that these results are based on the approximation~\eqref{Happ}. In turn, 
%interactions do not influence the steady-state expectation values $\langle {n}_k\rangle, \langle {n}_k {n}_q\rangle$, etc, and their impact on the current is only captured by the second term in Eq.~\eqref{Jss}, which gives a linear dependence on the interaction strength $U$. A more precise study on the effect of interactions can be performed in experiments. We will discuss the experimental implementation of our scheme in Section~\ref{sec:EI}.

Now let us turn on interactions. %The impact of interactions is not expected to be captured within the mean-field theory. 
For a fixed particle number, which can be calculated by using exact diagonalization~[the orange squares in Fig.~\ref{MF1}(a)], the interactions are found to have some impact on the steady-state heat current for small systems. While this effect becomes weaker with increasing system sizes, and thus does not change the scaling of the current with system size.
For a fixed filling factor, the numerical simulation is challenging. We resort to quantum jump Monte Carlo method~\cite{PhysRevLett.68.580,Daley2014Review}, which offers an efficient stochastic simulation of
the master equation by means of quantum trajectories. 
We are able to calculate the heat current of the interacting system for up to $20$ lattice sites~[see orange diamonds in Fig.~\ref{MF1}(b)]. We can clearly observe that it is reduced with respect to the heat current for the non-interacting system. Moreover, we can see that it drops with $M$. The accessible system sizes of $20$ do, however, not allow to reach the regime, where the ballistic transport of the non-interacting system becomes apparent from the saturation of the heat current. While it would have been interesting to study numerically, whether/how ballistic transport breaks down with increasing interactions, we would like to point out that our scheme opens the opportunity to investigate this question experimentally in a quantum simulation, where heat-current-carrying states also of larger interacting systems are prepared using feedback control.

\subsection{Influence of disorder}\label{sec:disorder}
\begin{figure}
	\centering
	%\includegraphics[width=0.9\columnwidth]{figure/MC_N2_M35_U0.pdf}\\
	%	\includegraphics[width=0.9\columnwidth]{figure/dim_log.pdf}
	%\includegraphics[width=1.9\columnwidth]{figure/Disorder_N3_M6.pdf}
	\includegraphics[width=0.9\columnwidth]{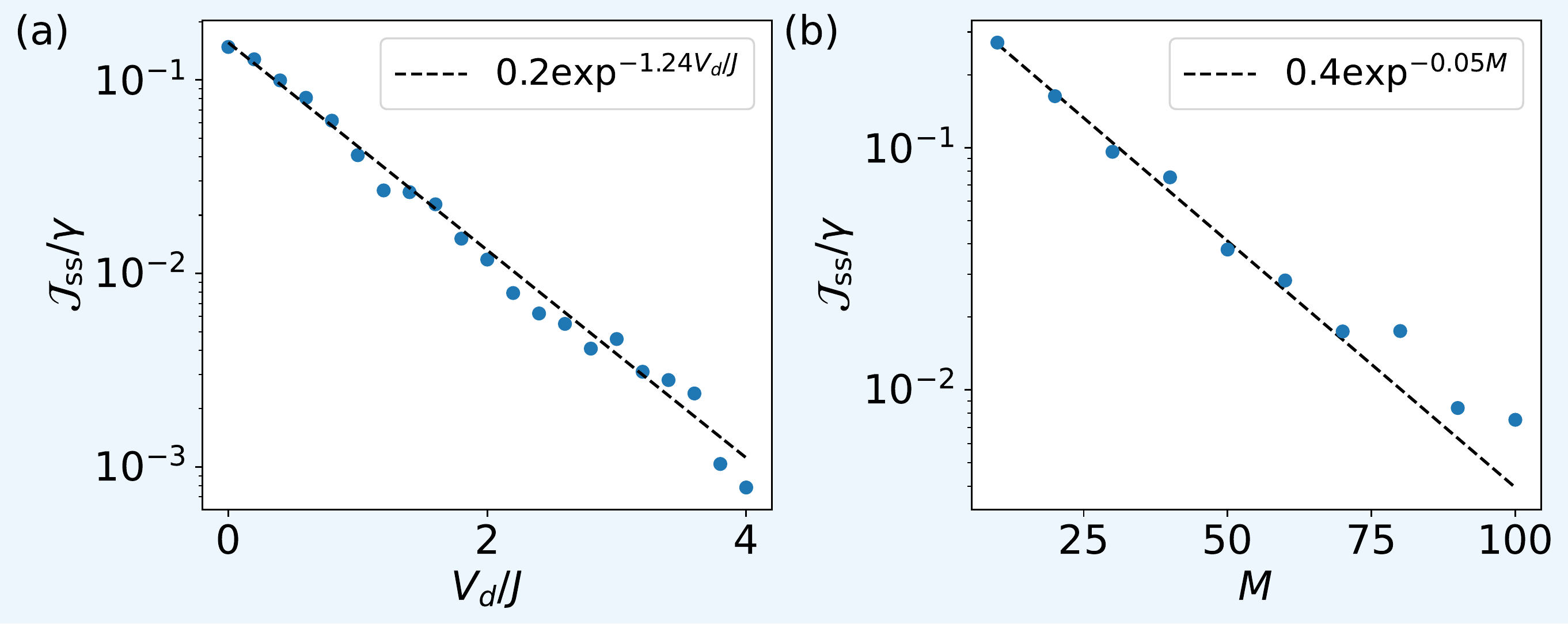}\\
	\includegraphics[width=0.9\columnwidth]{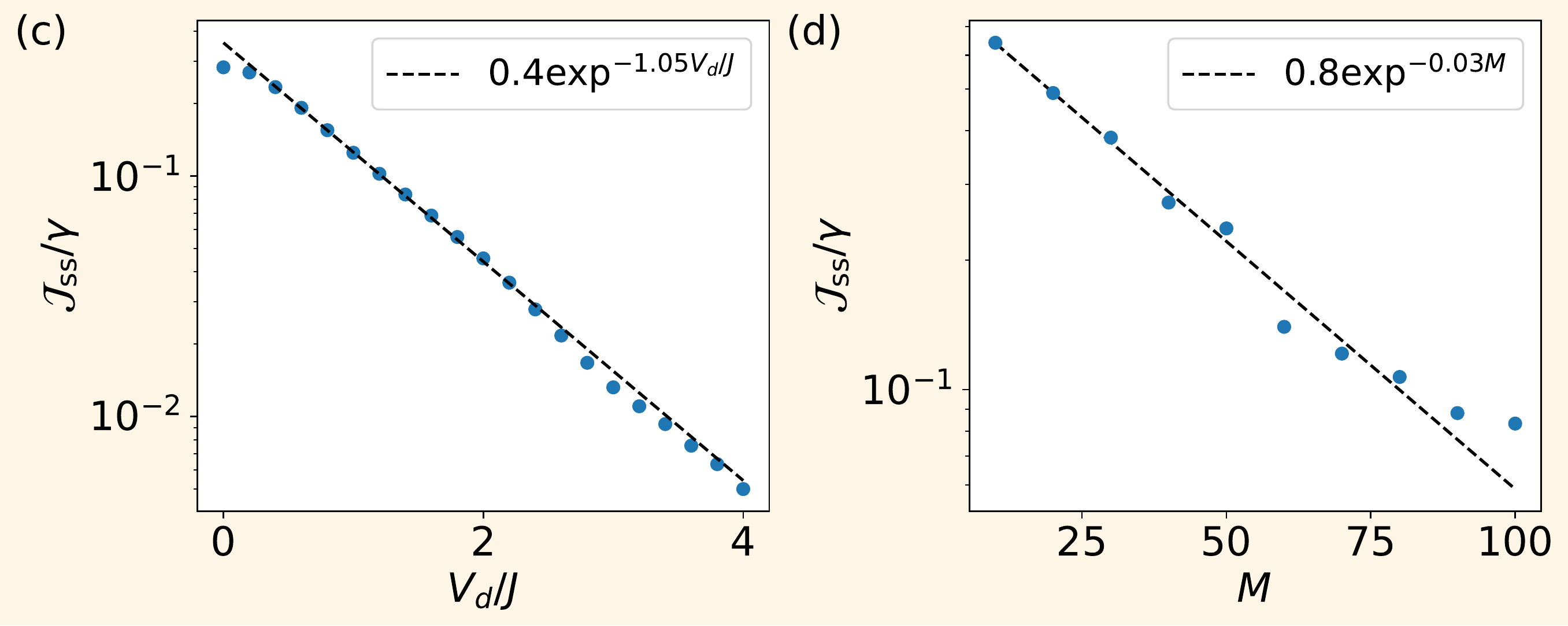}
	\caption{(a, c) The steady-state heat current ${\cal J}_{\rm ss}$ as a function of the disorder strength for one particle on $M=10$ sites. The results are obtained from exact diagonalization. (b, d) The system-size scaling of ${\cal J}_{\rm ss}$ at $V_d=0.2J$ for $N=M/5$. The results are obtained from kinetic theory with mean-field approximation.
		The panels in blue background (a, b) are the results for our scheme. The panels in yellow background (c, d) are the results when the system is coupled to thermal bath. All the results are averaged over $100$ trajectories with different disorder configurations. Parameters are $U=0$, $\gamma=0.01J$, $\lambda_L=0.1$, $\lambda_R=1$, $T_L=10J$, and $T_R=0.01J$.}
	\label{Disorder}
\end{figure}

Here we investigate the influence of disorder on the steady-state heat current. For this purpose,
%Now we add disorder to the system and see how it influences the steady state current. 
%Here we consider three different models.
we add a random on-site potential, with $V_l$ being a random number uniformly distributed in the range $[-V_d,V_d]$ . The results are shown in Fig.~\ref{Disorder}, which are averaged over $100$ trajectories with different disorder configurations. As expected, the current decreases with the disorder strength $V_d$, as shown in Fig.~\ref{Disorder}(a).  Figure~\ref{Disorder}(b) shows the current as a function of the lattice site number $M$ at $V_d=0.2J$ and fixed filling $N/M=1/5$. The decay of the current with increasing system size is approximately exponential, as indicated by the close-to-linear form of the current in the log-$y$ plot.
Similar behaviors are observed when the system is coupled to real thermal baths, as shown in Fig.~\ref{Disorder}(c) and (d).
%\begin{itemize}
%	\item {\bf Aubry-Andr{\'e} (AA) model}. We add quasi disorder via an incommensurate periodic potential modulation)
%	$V(l)=V_d\cos(2\pi\beta l)$, with $\beta=(\sqrt{5}-1)/2$.
%	\item {\bf Stark model}. We add a linear on-site potential, $V(l)=V_d l$.
%	\item {\bf Disorder model}. We add a random on-site potential, $V(l)=V_d x_l$, with $x_l$ being a random number uniformly distributed in the range $[-1,1]$. The results are averaged over $100$ trajectories with different disorder configurations.
%\end{{itemize}

%\section{Impact of interaction}

\section{Experimental Implementation}\label{sec:EI}
Now we discuss the experimental implementation of our scheme.
For the engineering of the local baths, one needs to perform measurements of the on-site population, and add the corresponding feedback control. The former can be implemented via homodyne detection of the off-resonant scattering of structured probe light from the atoms~\cite{PhysRevLett.114.113604,PhysRevLett.115.095301,RevModPhys.85.553}. To engineer two independent baths, one can for instance use two probe beams with different frequencies.  
The feedback control of tunneling with complex rate can be realized by modulating the on-site energy of the relevant sites~(see~Appendix~\ref{append:feedback}).
\begin{figure}
	\centering
	\includegraphics[width=0.9\columnwidth]{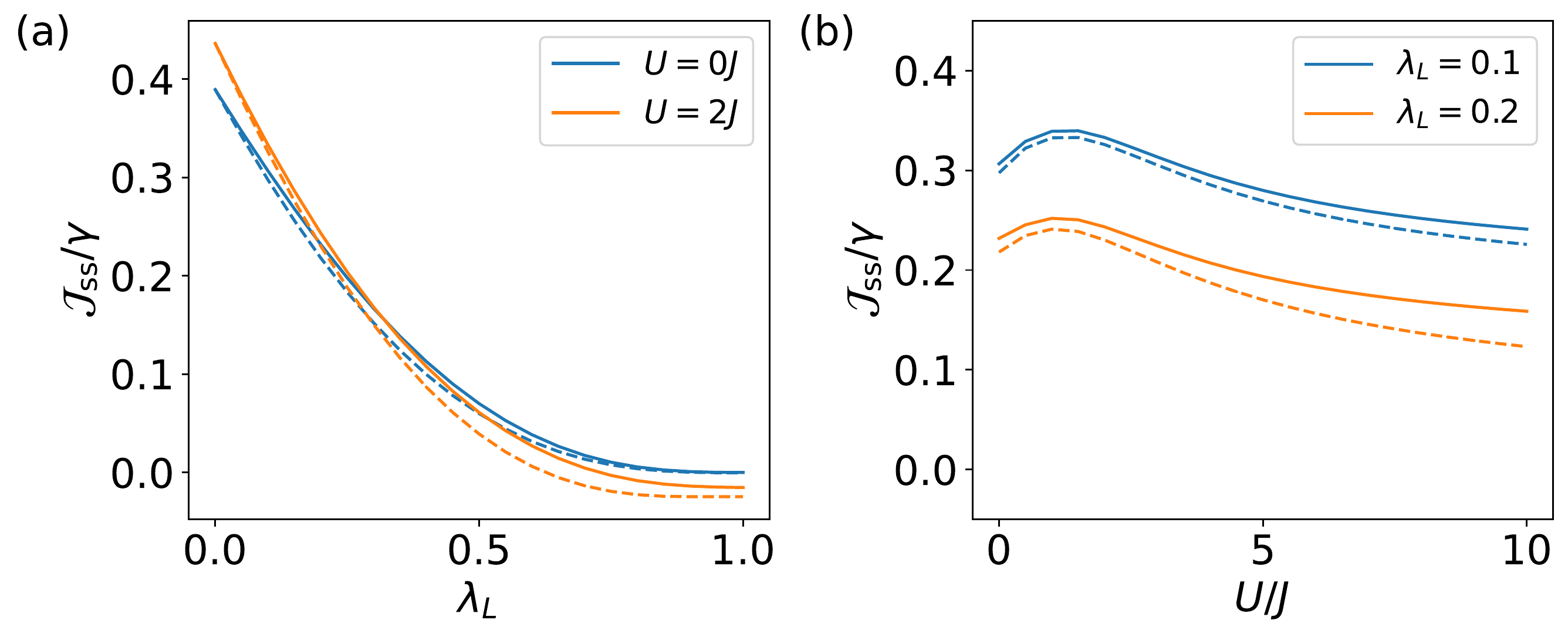}
	\caption{The steady-state heat current ${\cal J}_{\rm ss}$ as a function of (a) the feedback strength $\lambda_L$ for two different interaction strengths $U$, and (b) the interaction strength $U$ for two different feedback strength $\lambda_L$. Parameters are $N=2$, $M=10$, $V_l=0$, $\gamma=0.01J$, and $\lambda_R=1$. The solid lines are exact results. The dashed lines are the approximated results by using Wick's decomposition, Eqs.\eqref{app1}-\eqref{app2}.}
	\label{interaction}
\end{figure}

For the heat current, we propose to make use of the measurement of single-particle density matrices $\langle a_i^\dag a_j\rangle$, which is experimentally accessible~\cite{PhysRevLett.121.260401}.
It can be shown that the heat current is given by
\sloppy	

\begin{equation}\label{JL}
	\begin{aligned}
		\frac{{\cal J}_L}{\gamma } &= \frac{J}{2}[(x+x^{-1})^2+4\lambda_L^2]\langle a_1^\dag a_2 + a_2^\dag a_1\rangle %\notag\\
		+ \frac{J}{2}(\lambda_L^2+x^{-2})\langle a_2^\dag a_3 + a_3^\dag a_2\rangle \\
		& + J\lambda_L(x + x^{-1})\langle (a_1^\dag a_2 - a_2^\dag a_1)^2 \rangle %\notag\\
		-2J\lambda_L \langle (n_1-n_2)(x n_1 - x^{-1} n_2)\rangle \\
		& -\frac{J\lambda_L}{2} \langle (a_1^\dag a_3 + a_3^\dag a_1)(x n_1 - x^{-1} n_2) + {\rm h.c.}\rangle \\
		&-\frac{J\lambda_L}{2x}\langle(a_1^\dag a_2-a_2^\dag a_1)(a_2^\dag a_3-a_3^\dag a_2) + {\rm h.c.}\rangle \\
		&- \frac{U\lambda_L}{4}\langle (x n_1 - x^{-1}n_2)(n_1 - n_2)(a_1^\dag a_2 + a_2^\dag a_1) + {\rm h.c.}\rangle \\
		&- \frac{U\lambda_L}{4}\langle (n_1-n_2)(a_1^\dag a_2 + a_2^\dag a_1)(x n_1 - x^{-1}n_2)+ {\rm h.c.}\rangle \\
		&- U\lambda_L^2(\langle (n_1 - n_2)^2 \rangle- \langle(a_1^\dag a_2 + a_2^\dag a_1)^2\rangle).
	\end{aligned}
\end{equation}		
%\begin{equation}\label{JL}
%	\begin{aligned}
%	\frac{{\cal J}_L}{\gamma } &= \frac{J}{2}[(x+x^{-1})^2+4\lambda_L^2]\langle a_1^\dag a_2 + a_2^\dag a_1\rangle %\notag\\
%	+ \frac{J}{2}(\lambda_L^2+x^{-2})\langle a_2^\dag a_3 + a_3^\dag a_2\rangle \\
%	& + J\lambda_L(x + x^{-1})\langle (a_1^\dag a_2 - a_2^\dag a_1)^2 \rangle %\notag\\
%	-2J\lambda_L \langle (n_1-n_2)(x n_1 - x^{-1} n_2)\rangle \\
%	& -\frac{J\lambda_L}{2} \langle (a_1^\dag a_3 + a_3^\dag a_1)(x n_1 - x^{-1} n_2) + {\rm h.c.}\rangle %\notag\\
%	-\frac{J\lambda_L}{2x}\langle(a_1^\dag a_2-a_2^\dag a_1)(a_2^\dag a_3-a_3^\dag a_2) + {\rm h.c.}\rangle \\
%	&- \frac{U\lambda_L}{4}(\langle (x n_1 - x^{-1}n_2)(n_1 - n_2)(a_1^\dag a_2 + a_2^\dag a_1) + (n_1-n_2)(a_1^\dag a_2 + a_2^\dag a_1)(x n_1 - x^{-1}n_2)+ {\rm h.c.}\rangle) \\
%	&- U\lambda_L^2(\langle (n_1 - n_2)^2 \rangle- \langle(a_1^\dag a_2 + a_2^\dag a_1)^2\rangle).
%	\end{alighed}
%\end{equation}
By using Wick's theorem, corresponding to the mean-field approximation used already for the kinetic theory,
\begin{equation}
	\langle a_k^\dag a_q a_p^\dag a_l \rangle \approx \langle a_k^\dag a_q \rangle \langle a_p^\dag a_l\rangle + \langle a_k^\dag a_l\rangle \langle a_q a_p^\dag\rangle, \notag
\end{equation}
we have
\begin{equation}\label{app1}
	\begin{aligned}
	\langle (a_1^\dag a_2 - a_2^\dag a_1)^2 \rangle &\approx \langle a_1^\dag a_2 - a_2^\dag a_1 \rangle^2 + \langle a_1^\dag a_2\rangle^2 + \langle a_2^\dag a_1\rangle^2 
	- 2\langle n_1\rangle \langle n_2\rangle - \langle n_1\rangle - \langle n_2\rangle, 
	%		&=& -2|\langle a_1^\dag a_2 \rangle|^2+ 2\langle n_1\rangle \langle n_2\rangle + \langle n_1\rangle + \langle n_2\rangle, 
	\end{aligned}
\end{equation}
\begin{equation}
	\begin{aligned}
	&\langle (n_1-n_2)(x n_1 - x^{-1} n_2)\rangle \\
	&\approx x\langle n_1\rangle(2\langle n_1\rangle -\langle n_2\rangle + 1) + x^{-1}\langle n_2\rangle(2\langle n_2\rangle - \langle n_1\rangle + 1) - (x+x^{-1})|\langle a_1^\dag a_2\rangle|^2, 
	\end{aligned}
\end{equation}
\begin{equation}
	\begin{aligned}
	&\langle (a_1^\dag a_3 + a_3^\dag a_1)(x n_1 - x^{-1} n_2) + {\rm h.c.}\rangle  
	\approx\langle a_1^\dag a_3 + a_3^\dag a_1 \rangle (4x\langle n_1\rangle - 2x^{-1}\langle n_2\rangle + x) \\
	&- 4x^{-1}{\rm Re}(\langle a_1^\dag a_2\rangle \langle a_2^\dag a_3\rangle),
	\end{aligned}
\end{equation}
\begin{equation}\label{app2}
	\begin{aligned}
	&\langle(a_1^\dag a_2-a_2^\dag a_1)(a_2^\dag a_3-a_3^\dag a_2) + {\rm h.c.}\rangle \\
	&\approx 2\langle a_1^\dag a_2-a_2^\dag a_1\rangle\langle a_2^\dag a_3-a_3^\dag a_2 \rangle  + \langle a_1^\dag a_3 + a_3^\dag a_1\rangle (2\langle n_2\rangle + 1) - 4{\rm Re}(\langle a_1^\dag a_2\rangle \langle a_3^\dag a_2\rangle),
	\end{aligned}
\end{equation}
\begin{equation}
\begin{aligned}
	&\langle (n_1 - n_2)^2 \rangle- \langle(a_1^\dag a_2 + a_2^\dag a_1)^2\rangle \approx 2\langle n_1 - n_2\rangle^2 - 4{{\rm Re}(\langle a_1^\dag a_2\rangle^2) - 4{|\langle a_1^\dag a_2\rangle|}^2},
\end{aligned}	
\end{equation}
	\begin{equation}\label{app3}
		\begin{aligned}
		&\langle (x n_1 - x^{-1}n_2)(n_1 - n_2)(a_1^\dag a_2 + a_2^\dag a_1) + (n_1-n_2)(a_1^\dag a_2 + a_2^\dag a_1)(x n_1 - x^{-1}n_2)+ {\rm h.c.}\rangle \\
		& \approx 4 \langle a_1^\dag a_2 + a_2^\dag a_1\rangle\{x\left[6\langle n_1\rangle^2+4\langle n_1\rangle\right]+x^{-1}\left[6\langle n_2\rangle^2+4\langle n_2\rangle\right]\\
		&-(x+x^{-1})[4\langle n_1\rangle \langle n_2\rangle + \langle n_1\rangle + \langle n_2\rangle + 2|\langle a_1^\dag a_2\rangle|^2]\}. 
		\end{aligned}
	\end{equation}	
%			\end{widetext}
A comparison between the exact results of the individual terms and the mean-field approximation is presented in  ~\ref{append:check}.
In Fig.~\ref{interaction}, we compare the exact results of the steady-state heat current and the approximated ones by using Eqs.\eqref{app1}-\eqref{app2}. Note that for the latter, we neglect the interaction terms proportional to $U$ in Eq.~\eqref{JL} since they are small and cannot be expected to be captured within mean-field theory~(see~Appendix~\ref{append:check} for details). For the non-interacting case~[see the blue data in Fig.~\ref{interaction}(a)], the approximation is found to be very good, especially for $\lambda_L$ close to $0$ or $1$. For the interacting case~[see the orange data in Fig.~\ref{interaction}(a) and the results in Fig.~\ref{interaction}(b)], the approximated results still capture the  behavior very well. These results confirm the feasibility to measure the heat current for our scheme in experiments.  
%The single-particle density matrix can be measured in experiments~\cite{PhysRevLett.121.260401}.

\section{Conclusion}\label{sec:conclusion}
In conclusion, we have proposed a scheme for the realization of heat-current-carrying states of ultracold atoms in an optical lattice using Markovian feedback control.  Measurements and feedback control are implemented at the boundaries of the lattice to mimic the effect of coupling the system locally to two thermal baths with different temperature. We studied the  scaling of the steady-state heat current with system size by using two approaches: semi-classical Monte Carlo simulation and kinetic theory. For the non-interacting case, both approaches show good agreement with the  results from exact diagonalization (accessible for small systems). When the particle number is fixed, the current  decays with the lattice size as $M^{-1}$. For a fixed filling factor, the current is found to decay at first, but rapidly saturate at a finite value, independent of the system size. Namely, the system exhibits ballistic transport. 
%Interactions show impact on the current, but do not change its scaling behavior when the particle number is fixed.
For the interacting systems with a fixed filling factor, our simulations are restricted to rather small system sizes, so that it is hard to investigate, how ballistic transport is modified or destroyed as a result of interactions. However, our scheme opens a door towards the experimental investigation of this problem in a quantum simulator of ultracold atoms.  In the presence of disorder, the current for a system with a fixed filling factor is found to decay exponentially with the system size.
These results confirm that the heat current generated by the feedback-engineered baths shows the same scaling behavior as those resulting from actual thermal baths.  We also discussed the experimental implementation of our scheme and, in particular, described how the heat current can be measured in the laboratory. Our findings can be tested by available experimental techniques. Our approach opens a new path for the experimental investigation of  heat-current-carrying states of large interacting systems for which a theoretical prediction is challenging. Thus it offers a new route for the quantum simulation of transport phenomena with ultracold atoms.

\section*{Acknowledgements}
This research was funded by the German Research Foundation (DFG)
within the collaborative research center (SFB) 910 under project number 163436311.

\setcounter{figure}{0}
\setcounter{table}{0}
\renewcommand{\theequation}{A\arabic{equation}}
\renewcommand{\thefigure}{A\arabic{figure}}

\appendix

%\section{}

%\label{append:prove}
%Here we show that for non-interacting particles, with $\lambda=1$, $A_l |g\rangle^{\otimes N} = 0$, where $|g\rangle^{\otimes N}$ denotes the ground state of the system, with all particles occupying the single-particle ground state, $|{g}\rangle$. 
%
%Firstly, we show that this holds for a single particle. In this case, the collapse operator reduces to 
%\begin{equation}
%	A_l = \frac{g_{l+1}}{g_l}{|l\rangle\langle l|} - \frac{g_{l}}{g_{l+1}}{|l+1\rangle\langle l+1|} - (|l\rangle\langle l+1|-|l+1\rangle\langle l|).
%\end{equation}
%Applying it to the ground state $|g\rangle = \sum\nolimits_{l'} {g_{l'} |{l'}\rangle}$, we get
%\begin{equation}
%	A_l|g\rangle = g_{l+1}|l\rangle - g_l|l+1\rangle-g_{l+1}|l\rangle+g_l|l+1\rangle=0.
%\end{equation}
%
%When there are no interactions between the particles, the multi-particle problem is equivalent to the single-particle problem.

\section{}

\label{append:feedback}
Here we discuss the implementation of the feedback control terms. By including the feedback terms to the Hamiltonian, we arrive at
\begin{equation}\label{Htar}
	\begin{aligned}
		H'(t) &= -(J_L(t) a_1^\dag a_2 + J_R(t) a_{M-2}^\dag a_{M-1}+J \sum\limits_{l\neq 1, M-2}{a_l^\dag a_{l+1}}+ {\rm h.c.}) + H_U + H_V,
	\end{aligned}
\end{equation}
where $H_U$ and $H_V$ denote the original interaction and on-site potential terms in~\eqref{BH} and 
\begin{equation}
	J_\mu(t) = J + i\sqrt{\gamma}\lambda_\mu  I_{\rm hom}(t) =\sqrt{J^2+\gamma\lambda_\mu^2 I_{\rm hom}^2} e^{i\theta_\mu(t)}, \quad \lambda = L, R,
\end{equation}
with $\tan\theta_\mu = \sqrt{\gamma}\lambda_\mu I_{\rm hom}/J$.
Our goal is to implement such a Hamiltonian. 

We can achieve it by modulating the on-site energy of the relevant sites so that the system Hamiltonian reads
\begin{equation}
	H_{d}(t) = -\sum\limits_{l=1}^{M-1}{(J_l a_l^\dag a_{l+1} + {\rm h.c.})}+H_U + H_V + \Delta_L(t)n_1 - \Delta_R(t) (n_{M-1}+n_{M}).
\end{equation}
In the rotating frame with transformation~$U(t)=\exp\{i\int{[\Delta_L(t_1)n_1-\Delta_R(t_1)(n_{M-1}+n_{M})]dt_1}\},$ the Hamiltonian is given by
\begin{equation}\label{Hd}
	\begin{aligned}
		\tilde{H}_{d}(t) &= U H U^\dag + i \dot U U^\dag \\
		&= -(J_1e^{i\theta_L(t)} a_1^\dag a_2 + J_{M-2}e^{i\theta_R(t)} a_{M-2}^\dag a_{M-1}+ \sum\limits_{l\neq 1,M-2}{J_la_l^\dag a_{l+1}}+ {\rm h.c.}) + H_U + H_V,
	\end{aligned}
\end{equation}
with  $\theta_\mu(t)=\int \Delta_\mu(t_1) dt_1$. By comparing Eqs.~\eqref{Htar} and~\eqref{Hd}, one can read off 
\begin{equation}
	J_1 = \sqrt{J^2+\gamma\lambda_L^2 I_{\rm hom}^2}, J_{M-2} = \sqrt{J^2+\gamma\lambda_R^2 I_{\rm hom}^2}, \int \Delta_\mu(t_1) dt_1=\arctan(\sqrt{\gamma}\lambda_\mu I_{\rm hom}/J),
\end{equation}
and $J_l = J$  for $l\neq 1, M-2$.

\section{}

\label{append:check}

\begin{figure} 
	\centering
	\includegraphics[width=0.99\columnwidth]{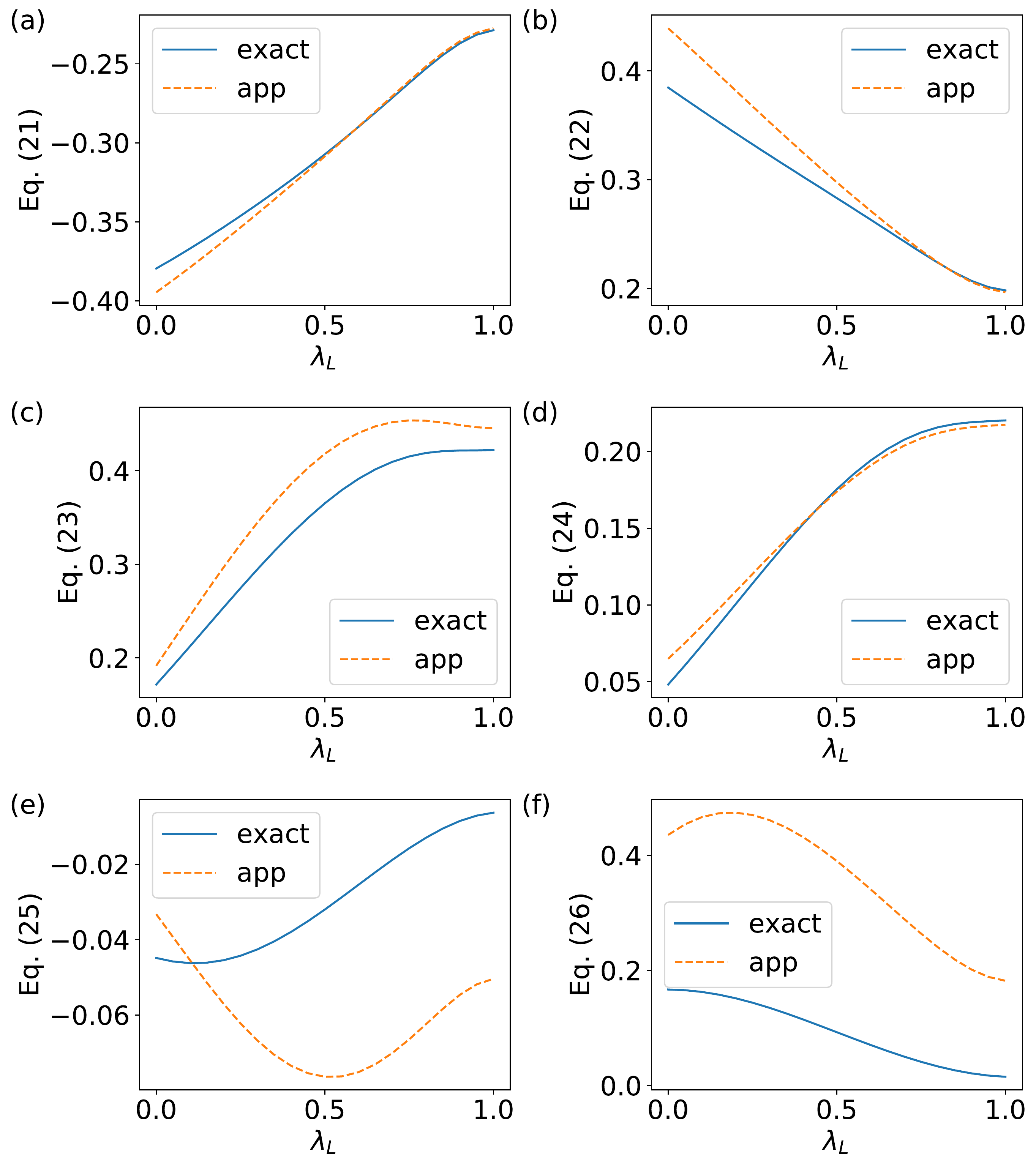}
	
	\caption{{Checking Eqs.~\eqref{app1}-\eqref{app3} in the main text. The solid~(dashed) lines denote the results of the left~(right) hand side of these equations. The parameters are $N=2$, $M=10$, $U=2J$, $V_l=0$, $\gamma=0.01J$, and $\lambda_R=1$.} 
	}\label{check}
\end{figure}

Here we check the validity of Eqs.~\eqref{app1}-\eqref{app3} in the main text. Figure~\ref{check} compares the left~(solid lines) and right~(dashed lines) hand side of these terms. One can see that for the terms relevant for tunneling effects, i.e., (a)-(d), the approximations are good. They become worse when it comes to the terms relevant to interactions, i.e., (e)-(f). The worse performance of the approximation in (f) is attributed to the involved higher order correlations compared with other terms. Due to the bad performance of the approximation in the two interaction-relevant terms, we neglect them in the calculation of the approximated heat current. Note that the term in (e) has very small value, and thus its influence is small. For the term in (f), from the expression of the heat current, Eq.~\eqref{JL} in the main text, one can see that it is proportional to $\lambda_L^2$, and thus its effect is weak for small $\lambda_L$. This observation is in consistent with the results shown in Fig.~\ref{interaction} of the main text.

\newpage

\section*{References}

\bibliography{ref}{}

\end{document}